\documentclass[12pt]{article}
\usepackage{amssymb,latexsym}
\usepackage[dvips]{graphicx}
\newtheorem{prop}{Proposition}
\newtheorem{conc}{Conclusion}

\def\beq{\begin{equation}}
\def\eeq{\end{equation}}
\def\bea{\begin{eqnarray}}
\def\eea{\end{eqnarray}}
\def\beas{\begin{eqnarray*}}
\def\eeas{\end{eqnarray*}}
\def\nn{\nonumber}
\def\ds{\displaystyle}
\def\[{[\![}
\def\]{]\!]}

\def\a{\alpha}
\def\b{\beta}
\def\g{\gamma}
\def\d{\delta}
\def\t{\theta}
\def\T{\Theta}
\def\P{{\bf P}}
\def\R{{\bf R}}

\def\M{{\bf M}}

\def\ra{\rangle}
\def\la{\langle}

\def\n{\noindent}
\def\N{\Phi}
\def\s{\smallskip}
\def\-{$\bullet\;$}
\def\ov{\overline}
\def\ds{\displaystyle}
\def\dag{\dagger}
\def\defn{\equiv}
\def\|{\,|\,}
\def\-{\!-\!}
\def\hR{\hat{R}}
\def\hP{\hat{P}}

\def\T{\Theta}
\def\hR{\hat{R}}
\def\hr{\hat{r}}
\def\hd{\hat{d}}
\def\hM{\hat{M}}
\def\hbR{\hat{\bf R}}
\def\hbP{\hat{\bf P}}
\def\hbr{\hat{\bf r}}
\def\ov{\overline}

\def\s{s}
\def\hs{\hat{s}}
\def\hbs{\hat{\bf s}}
\def\ph+{{\phantom{+}}}
\def\mod{{\rm mod\,}}

\begin{document}
%
%
\begin{center}
{\Large \bf
A non-commutative ${\bf n}$-particle 3D Wigner quantum oscillator}\\
[5mm]
{\bf R.C.~King}\footnote{E-mail: R.C.King@maths.soton.ac.uk}\\[1mm]
School of Mathematics, University of Southampton,\\
Southampton SO17 1BJ, U.K.;\\[2mm]
{\bf T.D.\ Palev}\footnote{E-mail: tpalev@inrne.bas.bg}\\[1mm]
Institute for Nuclear Research and Nuclear Energy,\\
Boul. Tsarigradsko Chaussee 72, 1784 Sofia, Bulgaria;\\[2mm]
{\bf N.I.~Stoilova}~\footnote{E-mail: Neli.Stoilova@UGent.be; Permanent address:
Institute for Nuclear Research and Nuclear Energy, Boul.\ Tsarigradsko Chaussee 72,
1784 Sofia, Bulgaria} {\bf and J.\ Van der Jeugt}\footnote{E-mail:
Joris.VanderJeugt@UGent.be}\\[1mm]
Department of Applied Mathematics and Computer Science,
University of Ghent,\\
Krijgslaan 281-S9, B-9000 Gent, Belgium.
\end{center}

\vskip 5mm
\begin{abstract}
\noindent
An $n$-particle 3-dimensional Wigner quantum oscillator model is constructed explicitly.
It is non-canonical in that the usual coordinate and linear momentum commutation relations
are abandoned in favour of Wigner's suggestion that Hamilton's equations and the Heisenberg
equations are identical as operator equations. The construction is based on the use of Fock
states corresponding to a family of irreducible representations of the Lie superalgebra
$sl(1|3n)$ indexed by an $A$-superstatistics parameter $p$. These representations are typical
for $p\geq3n$ but atypical for $p<3n$. The branching rules for the restriction from $sl(1|3n)$
to $gl(1)\oplus so(3)\oplus sl(n)$ are used to enumerate energy and angular momentum eigenstates.
These are constructed explicitly and tabulated for $n\leq2$.
It is shown that measurements of the coordinates of the individual particles gives rise
to a set of discrete values defining nests in the 3-dimensional configuration space. The fact
that the underlying geometry is non-commutative is shown to have a significant impact on
measurements of particle separation. In the atypical case, exclusion phenomena are identified
that are entirely due to the effect of
$A$-superstatistics. The energy spectrum and associated degeneracies are calculated for an
infinite-dimensional realisation of the Wigner quantum oscillator model obtained by summing
over all $p$. The results are compared with those applying to the analogous canonical quantum
oscillator.
\end{abstract}

\vfill\eject

\section{Introduction}
\setcounter{equation}{0} In a previous paper~\cite{JPhys}, hereafter referred to as
{\tt I}, we studied the properties of a non-canonical single particle 3-dimensional
Wigner quantum oscillator (WQO). In this model physical observables, such as energy,
angular momentum, position and linear momentum, were all associated with Hermitian
operators acting in a Hilbert space spanned by certain Fock
states~\cite{JPhys}-\cite{WQSJMP} arising in the representation theory of the Lie
superalgebra $sl(1|3)$ initiated by Kac~\cite{Kac}. This superalgebra was shown to
arise in a natural way [1] when seeking algebraic solutions to the compatibility
conditions necessary to ensure that, as suggested by Wigner~\cite{Wigner}, Hamilton's
equations and the Heisenberg equations are identical as operator equations in the
relevant Hilbert space. It is this suggestion, postulate (P3) in {\tt I}, for a Wigner
quantum system that replaces the canonical coordinate and linear momentum commutation
relations of a more conventional quantum theory. This idea of Wigner has been studied
by several authors from different points of view. Of the recent publications we
mention~\cite{A}-\cite{H}.

Here we extend this study in {\tt I} to the case of a non-canonical many particle
3-dimensional Wigner quantum oscillator. This time, in the case of $n$-particles the physical
observables are all associated with Hermitian operators acting in a Hilbert space spanned
by Fock states arising in the representation theory of the Lie superalgebra $sl(1|3n)$.
The relevance of this superalgebra is established once again by looking for algebraic
solutions of the $n$-particle compatibility conditions that ensure that as operator
equations in the relevant Hilbert space, Hamilton's equations and the Heisenberg equations
are identical.

The Hamiltonian itself of the $n$-particle WQO takes the form \beq
\hat{H}=\sum_{\alpha=1}^{n} \Big({ \hbP_\alpha^2 \over 2m}
+{m\omega^2\over{2}}{\hbR}_\a^2 \Big). \label{Ham0} \eeq Just as in the single particle
$n=1$ case, the Hilbert spaces, $W(n,p)$, studied here in the $n$-particle case are
associated with certain finite-dimensional irreducible representations of $sl(1|3n)$
specified by some non-negative integer $p$. Each of the Hilbert spaces $W(n,p)$,
spanned by Fock states $\|p,\T\ra$, is generated from a unique cyclic vector, $\|0\ra$,
by the action of the generators, $A_{\a i}^\pm$ of $sl(1|3n)$. As in {\tt I} each
generator $A_{\a i}^\pm$ is a certain linear combination of Hermitian coordinate and
linear momentum operators, $\hat{R}_{\a i}(t)$ and $\hat{P}_{\a i}(t)$ respectively,
for the particle specified by $\a$. Again, as in {\tt I}, the action of these
generators is constrained by the conditions: \beq
    A_{\a i}^-\|0\ra=0, \quad
    A_{\a i}^-A_{\b j}^+\|0\ra=p\,\delta_{\a\b}\,\delta_{ij}\|0\ra,
\quad i,j=1,2,3,\quad \a,\b=1,2,\ldots,n. \label{Wnp} \eeq The resulting
finite-dimensional irreducible representations are special cases of those classified by
Kac~\cite{Kac} and include both typical and atypical
representations~\cite{sl(m/n)typ}-\cite{sl(m/n)atyp}  characterized here by $p\geq3n$
and $p<3n$, respectively.

Having set up this mathematical machinery for the WQO in Section~2, what follows is
discussion of a range of physical properties of this multi-particle system. The energy
spectrum and angular momentum are discussed in Section~3. This leans heavily on
developments made in a sequence of previous papers~\cite{JPhys}-\cite{WQSJMP},
\cite{Palev3}-\cite{Palev4}. Here the complications associated with the determination
of energy and angular momentum eigenstates in the many particle context are dealt with
by the use of the branching rules appropriate to the restrictions, first from
$sl(1|3n)$ to $gl(1)\oplus sl(3n)$~\cite{Dondi}-\cite{Palev5}   then from $sl(3n)$ to
$sl(3)\oplus sl(n)$~\cite{King3} and finally from $sl(3)$ to $so(3)$~\cite{King3}-
\cite{Hamermesh}, where $so(3)$ is the algebra associated with the total angular
momentum of the system. By way of illustration, energy and angular momentum eigenstates
are tabulated explicitly for both $n=1$ and $n=2$.

Section~4 is concerned with Wigner quantum oscillator configurations. It is found that
just as energy and angular momentum are discretely quantised, so are the coordinates,
however measured, of each of the $n$-particles. As observed previously in the
$1$-particle case analysed in {\tt I}, the eigenvalues $r_{\a k}$ of the position
operators $\hat{r}_{\a k}(t)$ of the $\a$th particle specify nests with coordinates
$\pm\sqrt{p-m}$, with the integer $m$ now taking on the values
$0,1,\ldots,\min(p,3n)-1$. The non-commutative nature of the geometry is such that once
again the position of any individual particle can not be specified precisely. That is
to say for each $\a$ there is no common eigenstate of $r_{\a k}(t)$ for all $k=1,2,3$.
This observation gives notice of the fact that  the interpretation of the measurement
of the distance between any two particles has to be undertaken with care. This is also
explored in Section~4 where it is shown that the expectation value of the square of the
separation distance in any of the stationary states $\|p,\T\ra$ can be interpreted as
the average of the square of the distance between the appropriate nests, weighted with
respect to the probabilities of occupying each nest.

In Section~5 it is shown that the $A$-superstatistics of this $sl(1|3n)$ multiparticle WQO
model leads to some exclusion phenomena whereby the state of one particle is influenced,
or even determined, by the states of the other particles even though the original
Hamiltonian~(\ref{Ham0}) is that of $n$ non-interacting particles. Finally a 3-dimensional
$n$-particle WQO model based on an infinite-dimensional Hilbert space
$W=\sum_{p=0}^\infty W(n,p)$ is compared and contrasted with an analogous canonical
quantum oscillator (CQO) model. Both the WQO and the CQO models are shown to involve equally
spaced energy levels, but their ground states, energy gaps and degeneracies are all shown to
differ markedly in the two models.

\section{Wigner quantum oscillators}
\setcounter{equation}{0}

Let $\hat{H}$ be the Hamiltonian of an $n$ particle three-dimensional
harmonic oscillator, that is
\beq
\hat{H}=\sum_{\alpha=1}^{n} \Big({ \hbP_\alpha^2 \over 2m}
+{m\omega^2\over{2}}{\hbR}_\a^2 \Big). \label{H}
\eeq
We proceed to view this oscillator as a Wigner quantum system.
According to postulate (P3) in {\tt I} the three-dimensional vector operators
$\hbR_1,\ldots,\hbR_n$ and $\hbP_1,\ldots,\hbP_n$ have to be defined in such a
way that Hamilton's equations
\beq
    {\dot{\hbP}}_\a=-m\omega^2\hbR_\a, \ \ {\dot{\hbR}}_\a = {1\over m}\hbP_\a
    \ ~ {\rm for} ~\ \a=1,2,\ldots,n,
     \label{Ham}
\eeq
and the Heisenberg equations
\beq
     {\dot{\hbP}}_\a = {i\over{\hbar}}[\hat{H},\hbP_\a], \ \
     {\dot{\hbR}}_\a = {i\over{\hbar}}[\hat{H},\hbR_\a]
     \ ~ {\rm for} ~ \ \a=1,2\ldots,n,
     \label{Heis}
\eeq
are identical as operator equations. These compatibility conditions are
such that
\beq
   [\hat{H},\hbP_\a]=i\hbar m \omega^2\hbR_\a ,\ \
   [\hat{H},\hbR_\a]=-{{i\hbar}\over{m}}\hbP_\a
    \ ~ {\rm for} ~ \ \a=1,2,\ldots,n.
     \label{comp}
\eeq
To make the connection with $sl(1|3n)$ we write the operators
$\hbP_\a$ and $\hbR_\a$
for $\a=1,2,\ldots,n$ in terms of new operators:

\beq
A_{\alpha k}^\pm= \sqrt{(3n-1)m \omega \over 4\hbar}
 \hR_{\a k} \pm i
  \sqrt {(3n-1)\over 4m \omega \hbar} \hP_{\a k}, ~~k=1,2,3.
 \label{A}
\eeq
The Hamiltonian $\hat{H}$ of~(\ref{H}), the single particle Hamiltonians,
$\hat{H}_\a$, and the
compatibility conditions~(\ref{comp}) take the form~\cite{WQSJMP}:
\beq
     \hat{H}=\sum_{\a=1}^n \hat{H}_\a \ \ \hbox{with}\ \
    \hat{H}_\a = {{\omega\hbar}\over{3n-1}}\sum_{i=1}^3 \{A_{\a i}^+,A_{\a i}^-\},
     \label{Halpha}
\eeq
\beq
\sum_{\b=1}^n  \sum_{j=1}^3  [ \{A_{\b j}^+,A_{\b j}^- \},A_{\a i}^\pm]
=\mp (3n-1)A_{\a i}^\pm , \quad i,j=1,2,3, \quad \a,\b =1,2,\ldots , n.
\label{comp1}
\eeq
As a solution to~(\ref{comp1}) we chose operators $A_{\a i}^\pm$ that satisfy the
following triple relations:
\bea
&& [\{A_{\a i}^+,A_{\b j}^-\},A_{\g k}^+]=
\delta_{jk}\delta_{\b \g}A_{\a i}^+
-\delta_{ij}\delta_{\a \b}A_{\g k}^+,  \nn\\
&& [\{A_{\a i}^+,A_{\b j}^-\},A_{\g k}^-]=
-\delta_{ik}\delta_{\a \g}A_{\b j}^-
+\delta_{ij}\delta_{\a \b}A_{\g k}^-, \label{sl}\\
&& \{A_{\a i}^+,A_{\b j}^+\}=
\{A_{\a i}^-,A_{\b j}^-\}=0. \nn
\eea

\begin{prop}
The operators $A_{\a i}^\pm$, for $i=1,2,3$ and
$\alpha=1,2,\ldots,n$,
are odd elements generating the Lie superalgebra
$sl(1|3n)$.
The operators $\{A_{\a i}^+,A_{\b j}^-\}$ for $i,j=1,2,3$ and
$\a,\b=1,2,\ldots,n$, are even elements generating the maximal
even Lie subalgebra $gl(3n)$.
\end{prop}

The Lie superalgebra is from class $A$ in the classification of
the basic classical Lie superalgebras~\cite{Kac}. As we have indicated,
the corresponding statistics is referred to as
$A$-superstatistics~\cite{PSJ}. The generators $A_{\a i}^\pm$ are said to be
creation and annihilation operators (CAOs) of $sl(1|3n)$. These
CAOs are the analogue of the Jacobsen generators for the Lie
algebra $sl(3n+1)$~\cite{A-stat} and could also be called Jacobsen
generators of $sl(1|3n)$.

We would underline the fact that all considerations here are in
the Heisenberg picture. The position and momentum operators
depend on time. Hence also the CAOs depend on time. Writing this
time dependence explicitly, one has:

\bea
& \hbox{Hamilton's  equations}&
{\dot A}_{\a k}^\pm(t)=\mp i \omega A_{\a k}^\pm(t), \label{Ham1} \\
& \hbox{Heisenberg  equations}&
{\dot A}_{\a k}^\pm(t)=-{{i \omega }\over{3n-1}}\sum_{\b=1}^n \sum_{j=1}^3
[A_{\a k}^\pm(t),\{A_{\b j}^+(t), A_{\b j}^-(t)\}]. \nn\\
&&\label{Heis1}
\eea
The solution of~(\ref{Ham1})  is evident,
\beq
A_{\a k}^\pm(t)= A_{\a k}^\pm(0)\ e^{\mp i\omega t} \label{sol}
\eeq
and therefore if the defining relations~(\ref{sl}) hold
at a certain time $t=0$, i.e., for $A_{\a k}^\pm \equiv A_{\a k}^\pm(0) $,
then they hold as equal time relations for any other time $t$. From~(\ref{sl})
it follows also that the Eqs.~(\ref{Ham1}) are identical with Eqs.~(\ref{Heis1}).
For further use we write the time dependence also of $\R_\a$
and $\P_\a$ explicitly:
\bea
&& \hR_{\a k}(t)={\sqrt{\hbar \over {(3n-1)m\omega} }} (A_{\a k}^+ e^{-i \omega
t}+A_{\a k}^- e^{i \omega t}), \label{Rt} \\
&& \hP_{\a k}(t)=-i\sqrt{m \omega \hbar
\over {(3n-1)}} (A_{\a k}^+ e^{-i \omega t}- A_{\a k}^-e^{i \omega
t}). \label{Pt}
\eea

\n
Finally, the single particle angular momentum operators
$\hM_{\a j}$ defined in~\cite{WQSJMP} by
\beq
      \hM_{\a j} = -{{3n-1}\over{2\hbar}} \sum_{k,l=1}^3
    \ \epsilon_{jkl} \{\hR_{\a k},\hP_{\a l}\}
      \ \ \a=1,2,\ldots,n,\ \ j=1,2,3,
      \label{MRP}
\eeq
take the following form:
\beq
      \hM_{\a j} = -i \sum_{k,l=1}^3 \epsilon_{jkl}
   \{ A_{\a k}^+,A_{\a l}^-\}, \quad j=1,2,3.
      \label{MA}
\eeq
In terms of these operators the three components of the total angular
momentum
operator $\hat{\M}$ are given by
\beq
\hM_j=\sum_{\a=1}^n \hM_{\a j},\quad j=1,2,3. \label{M}
\eeq
It is straightforward to verify that with respect to this choice
of angular momentum operator $\hat{\M}$ the operators $\hbR_\a$, $\hbP_\a$,
$\hat{\M}_\a$ and $\hat{\M}$ all transform as $3$-vectors.

As indicated in Proposition~1 the CAO's $A_{\a i}^{\pm}$ with
$\a=1,2,\ldots,n$ and $i=1,2,3$ generate the Lie superalgebra
$sl(1|3n)$. This superalgebra has both finite and
infinite-dimensional irreducible representations. Here we will
consider only finite-dimensional irreducible representations.
These have been classified by Kac~\cite{Kac} and are subdivided into
typical and atypical irreducible representations. The typical
irreducible representations coincide with the corresponding
Kac-modules~\cite{sl(m/n)typ} for which there exists a rather simple
character formula and a dimension formula. The same is not true
of the atypical irreducible representations. Their dimensions are
less than would be given by the dimensions of the corresponding
Kac-modules~\cite{sl(m/n)atyp}.

The representations of $sl(1|3n)$ that are of interest here are
those finite-dimensional covariant tensor irreducible
representations, $V^p$, with highest weight $(p,0,\ldots,0)$ for
some non-negative integer $p$. Such representations are typical if
$p\geq3n$ and atypical if $p<3n$, and have dimension given by:
\beq
   \dim V^p = \sum_{q=0}^{\min(p,3n)} \left({{3n}\atop{q}}\right)
            = \cases{
   {\ds{\sum_{q=0}^{p} \left({{3n}\atop{q}}\right)}}
        &if $V^p$ is atypical, i.e. $p<3n$;\cr\cr
     \ \ 2^{3n}&if $V^p$ is typical, i.e. $p\geq 3n$.\cr}
   \label{dim}
\eeq All of these irreducible representations, $V^p$, whether typical or atypical,  may
be constructed explicitly by means of the usual Fock space technique, as well as others
for which $p$ is not an integer. In our $A$-superstatistics case they may be
constructed, precisely as in the parastatistics case~\cite{Green}, from the requirement
that the corresponding representation space, $W(n,p)$, contains (up to a multiple) a
unique cyclic vector $\|0\ra$ such that \beq
    A_{\a i}^-\|0\ra=0, \quad
    A_{\a i}^-A_{\b j}^+\|0\ra=p\delta_{\a\b}\delta_{ij}\|0\ra,
\quad i,j=1,2,3,\quad \a,\b=1,2,\ldots,n.
  \label{defrepr}
\eeq
The above relations are enough for the construction of the full
representation space $W(n,p)$. This space defines an
indecomposable finite-dimensional
representation of the CAO's~(\ref{sl}) and hence of
$sl(1|3n)$ for any value of $p$. However we wish to impose the
further physical requirements that:
\smallskip
\begin{itemize}
\item[(a)] $W(n,p)$ is a Hilbert space with respect to the natural
Fock space inner product;
\item[(b)] the observables, in particular the position and momentum
operators~(\ref{Rt})-(\ref{Pt}), are Hermitian operators.
\end{itemize}
Condition (b) reduces to the requirement that the Hermitian conjugate of
$A_{\a i}^+$ should be $A_{\a i}^-$, i.e.
\beq
      (A_{\a i}^\pm)^\dag=A_{\a i}^\mp.   \label{conj}
\eeq
The condition (a) is then such that $p$ is restricted to be a
non-negative integer~\cite{Palev1}, in fact any non-negative integer. We then
refer to $p$ as the order of the statistics. As a consequence
 the representation space $W(n,p)$ is
irreducible (and finite-dimensional).
It provides a concrete realization of the irreducible representation
$V^p$ of $sl(1|3n),$ of dimension given by~(\ref{dim}), as follows.

Let
\beq
     \Theta\defn(\t_{11},\t_{12},\t_{13},\t_{21},\t_{22},\t_{23},\ldots ,
                  \t_{n1},\t_{n2},\t_{n3}).        \label{Theta}
\eeq The state space $W(n,p)$ of the system, corresponding to an order of statistics
$p$, is spanned by the following orthonormal basis (called the $\Theta$-basis): \bea
&&\|p;\T\ra\defn \|p;.\t_{\a 1},\t_{\a 2},\t_{\a 3}.\ra
\defn\|p;\t_{11},\t_{12},\t_{13},\t_{21},\t_{22},
\t_{23},\dots ,
                  \t_{n1},\t_{n2},\t_{n3}\ra  \nn \\
&&  ~~~~~~~~~=\sqrt{{{(p-q)!}\over{p!}}} (A_{11}^+)^{\t_{11}}
(A_{12}^+)^{\t_{12}}(A_{13}^+)^{\t_{13}}
      (A_{21}^+)^{\t_{21}} (A_{22}^+)^{\t_{22}} (A_{23}^+)^{\t_{23}} \cdots\nn \\
 &&~~~~~~~~~
 \times  (A_{n1}^+)^{\t_{n1}} (A_{n2}^+)^{\t_{n2}} (A_{n3}^+)^{\t_{n3}} \|0\ra,
 \label{basis}
\eea
where
\beq
 \t_{\a i}\in\{0,1\}\ \ \hbox{for all}\ \ \a=1,2,\ldots,n,\ i=1,2,3
    \label{theta}
\eeq
and
\beq
 q\defn\sum_{\a=1}^n\sum_{i=1}^3\t_{\a i}\ \ \hbox{with}\ \
 0\leq q\leq\min(p,3n). \label{q}
\eeq
The transformation of the basis states~(\ref{basis})  under the action of
the CAO's reads as follows:
\bea
&&    A_{\a i}^-\|p;\T\ra
=\t_{\a i}(-1)^{\psi_{\a i}}\sqrt{p-q+1}
\|p;\T\ra_{\ov{\a i}},\label{action-}\\
 &&   A_{\a i}^+\|p;\T\ra
=(1-\t_{\a i})(-1)^{\psi_{\a i}}\sqrt{p-q}
\|p;\T\ra_{\ov{\a i}},\label{action+}
\eea
where
\beq
    \psi_{\a i}=\sum_{(\b j)<(\a i)} \theta_{\b j},
     \label{psi}
\eeq
with the ordering on the pairs $(\a i)$ defined by~(\ref{basis}), so that
$(\b j)<(\a i)$ if and only if either $\b<\a$ or $\b=\a$ and
$j<i$, and $\|p;\T\ra_{\ov{\a i}}$ are the states obtained from
$\|p;\T\ra$ after the replacement of $\t_{\a i}$ by
$\ov{\t}_{\a i}=1-\t_{\a i}$.

In what follows we shall also require the explicit action
of the anticommutators $\{A_{\a i}^+,A_{\b j}^-\}$
on the states $\|p;\T\ra$. This is given by:
\beq
    \{A_{\a i}^+,A_{\b j}^-\} \|p;\T\ra
  =\cases{
(p-q+\t_{\a i})\|p;\T\ra&if $(\a i)=(\b j)$;\cr\cr
(-1)^{\psi_{\b j}-\psi_{\a i}}\t_{\b j}(1-\t_{\a i})
\|p;\T\ra_{{\ov{\a i}},{\ov{\b j}}}&if $(\a i)<(\b j)$;\cr\cr
-(-1)^{\psi_{\a i}-\psi_{\b j}}\t_{\b j}(1-\t_{\a i})
\|p;\T\ra_{{\ov{\a i}},{\ov{\b j}}}&if $(\a i)>(\b j)$.\cr}
  \label{A+A-}
\eeq Here $\|p;\T\ra_{{\ov{\a i}},{\ov{\b j}}}$ denotes the state obtained from
$\|p;\T\ra$ by replacing $\t_{\a i}$ and $\t_{\b j}$ with $\ov{\t}_{\a i}=1-\t_{\a i}$
and $\ov{\t}_{\b j}=1-\t_{\b j}$, respectively.

Returning to~(\ref{basis}),
note the first big difference between the non-canonical WQO and
the case of a conventional CQO: {\it each state space $W(n,p)$ of
our WQO is finite-dimensional}. In fact the dimension is easily
seen from~(\ref{basis})  to coincide with that given for $V^p$ by~(\ref{dim}).

On the other hand in the CQO case it is well known that the corresponding (bosonic)
Fock space is spanned by the states
\bea
&&\|\N\ra\defn
\|\phi_{11},\phi_{12},\phi_{13},\phi_{21}, \phi_{22},\phi_{23},\ldots,
   \phi_{n1},\phi_{n2},\phi_{n3}\ra
= \prod_{\a=1}^n \prod_{i=1}^3 {1\over{\sqrt{\phi_{\a i}!}}}\nn \\
&&\times (B_{11}^+)^{\phi_{11}} (B_{12}^+)^{\phi_{12}}
          (B_{13}^+)^{\phi_{13}}
          (B_{21}^+)^{\phi_{21}} (B_{22}^+)^{\phi_{22}} (B_{23}^+)^{\phi_{23}}
 \cdots   (B_{n1}^+)^{\phi_{n1}} (B_{n2}^+)^{\phi_{n2}} (B_{n3}^+)^{\phi_{n3}}
          \|0\ra \nn\\
&&           \label{Phi}
\eea
with
\beq
  \phi_{\a i}\in\{0,1,2,\ldots\}\ \ \hbox{for all}\ \ \a=1,2,\ldots,n,
\ \ i=1,2,3.
   \label{phi}
\eeq
This space is clearly infinite-dimensional. The action of the
bosonic operators on these states is given by:
\beq
      B_{\a i}^- \|\N\ra= \sqrt{\phi_{\a i}}\|\N\ra_{-\a i}
\ \ \hbox{and}\ \  B_{\a i}^+ \|\N\ra=\sqrt{\phi_{\a i}+1}\|\N\ra_{+\a i}
       \label{Bose}
\eeq
where $\|\N\ra_{\pm\a i}$ are the states obtained from $\|\N\ra$
by the replacement of $\phi_{\a i}$ by $\phi_{\a i}\pm 1$, and there
is no upper bound on $\phi_{\a i}$.

\section{Physical properties - energy spectrum and angular momentum}
\setcounter{equation}{0}

We now discuss some of the physical properties of the
Wigner quantum oscillator (WQO), comparing them with those of the
canonical quantum oscillator (CQO).

The first thing to note is that, as in the case of the CQO, the
physical observables $\hat{H}$, $\hat{H}_\a$, $\hbR_\a$, $\hbP_\a$, $\hat{\M}$, and
$\hat{\M}_\a$ for $\a=1,2,\ldots,n$ are, in the case of the WQO, all
Hermitian operators within every Hilbert space $W(n,p)$ for each
$p=0,1,\ldots$.

Secondly, in the case of the WQO the Hamiltonian $\hat{H}$ is diagonal in the
basis~(\ref{basis})-(\ref{q}), i.e. the basis vectors
$\|p;\T\ra$ are stationary states
of the system. In each Hilbert space $W(n,p)$ there is a finite number of equally
spaced energy levels, with spacing $\hbar\omega$: \beq
     \hat{H}\|p;\T\ra = E_q\|p;\T\ra\ \ \hbox{with}\ \
     E_q={\hbar\omega}\left({{3np}\over{3n-1}}-q\right)
     \ \ \hbox{for}\ q=0,1,2,\ldots,\min(3n,p).
    \label{Eq}
\eeq

Similarly, for the CQO the Hamiltonian $\hat{H}$ is diagonal in the
basis~(\ref{Phi})-(\ref{phi}), so that the basis vectors $\|\N\ra$ are stationary
states. Now however, there is an infinite number of equally
spaced energy levels, but with the same spacing $\hbar\omega$:
\beq
     \hat{H}\|\N\ra = E_q\|\N\ra \ \ \hbox{where}\ \
     E_q= {\hbar\omega}\left( {3\over2}n+q\right)\ \ \hbox{with}\ \
     \ \ q=\sum_{\a=1}^n\sum_{i=1}^3 \phi_{\a i}=0,1,2,\ldots
     \label{BEq}
\eeq
The fact that the energy spectrum of the WQO is as given in~(\ref{Eq})
can be seen by noting that under the restriction from the Lie
superalgebra $sl(1|3n)$ to its reductive Lie subalgebra
$gl(1)\oplus sl(3n)$ the representation $W(n,p)=V_{sl(1|3n)}^p$
decomposes in accordance with the branching rule~\cite{Dondi}-\cite{King1}:
\bea
&      sl(1|3n)&\longrightarrow gl(1)\oplus sl(3n)\nn \\
&     V_{sl(1|3n)}^p &\longrightarrow
\sum_{q=0}^{\min(p,3n)} V_{gl(1)}^{q+3n(p-q)} \otimes V_{sl(3n)}^{1^q},
    \label{branch}
\eea
where the subscripts on the representation labels indicate the
relevant Lie algebra or superalgebra, and the superscripts are
the highest weights of the representation written in partition
notation. Here the Hamiltonian is just $\hbar\omega/(3n-1)$ times
the generator, $\sum_{\a=1}^n\sum_{i=1}^3\{A_{\a i}^+,A_{\a
i}^-\}$, of $gl(1)$, so that its eigenvalues $E_q$ are precisely
as given in~(\ref{Eq}).

The branching rule~(\ref{branch})  gives some additional information. The
notation in~(\ref{branch})  is such that $1^q$ signifies the partition
$(1,1,\ldots,1)$ all of whose $q$ non-vanishing parts are $1$.
Thus $V_{sl(3n)}^{1^q}$ signifies the $q$th rank totally
antisymmetric covariant irreducible representation of $sl(3n)$.
The degeneracy of the equally spaced states of energy $E_q$ is
then given by
\beq
      \dim V_{sl(3n)}^{1^q} = \left({{3n}\atop{q}}\right),
      \label{dimsl}
\eeq
as required for consistency with~(\ref{dim}).

In exactly the same way the fact that the energy spectrum of the
CQO is as given by~(\ref{BEq})  can be seen by considering the
restriction from the Lie superalgebra $osp(1|6n)$ first to its
even Lie subalgebra $sp(6n)$, generated by $\{B_{\a i}^{\xi},B_{\b
j}^{\eta}\}$ for all $\a,\b=1,2,\ldots,n$, $i,j=1,2,3$ and $\xi, \eta =\pm$~\cite{Palev2},
and then to the reductive Lie subalgebra $gl(1)\oplus sl(3n)$. Let
the infinite-dimensional irreducible representation of
$osp(1|6n)$ spanned by the basis states $\|\N\ra$ given by~(\ref{Phi})-(\ref{phi})
be denoted by $V_{osp(1|6n)}^{\varepsilon}$, where $\varepsilon$
is the weight vector $({1\over2},{1\over2},\ldots,{1\over2})$ in
the relevant $3n$-dimensional weight space. This decomposes first
into the sum of the two infinite-dimensional irreducible
metaplectic or oscillator representations of
$sp(6n)$~\cite{Moshinsky}-\cite{King2}
 of $sp(6n)$, which we denote here
by $V_{sp(6n)}^{\varepsilon_\pm}$, and then into
finite-dimensional irreducible representations of $gl(1)\oplus
sl(3n)$, all in accordance with the following branching rules:
\bea
 && osp(1|6n)\longrightarrow sp(6n)
  \longrightarrow gl(1)\oplus sl(3n)\nn \\
 && V_{osp(1|6n)}^{\varepsilon}
 \longrightarrow
 V_{sp(6n)}^{\varepsilon_+} \oplus V_{sp(6n)}^{\varepsilon_-}
 \longrightarrow
 \sum_{q=0}^{\infty} V_{gl(1)}^{{3\over2}n+q} \otimes V_{sl(3n)}^{q}.
    \label{osp}
\eea
This time the Hamiltonian is $\hbar\omega$ times the generator,
${1\over2}\sum_{\a=1}^n\sum_{i=1}^3\{B_{\a i}^+,B_{\a i}^-\}$, of
$gl(1)$, so that its eigenvalues $E_q$ are as given in~(\ref{BEq}).

Once again~(\ref{osp})  carries additional information on degeneracies.
This time in~(\ref{osp})  the superscript $q$ signifies the one part
partition $(q,0,\ldots,0)$. Thus $V_{sl(3n)}^{q}$ signifies the
$q$th rank totally symmetric covariant irreducible representation
of $sl(3n)$. The degeneracy of the equally spaced states of
energy $E_q$ in this CQO case is then given by
\beq
      \dim V_{sl(3n)}^{q} = \left({{3n-1+q}\atop{q}}\right).
      \label{dimslq}
\eeq

This analysis shows that the energy levels of both the WQO and
the CQO are equally spaced, both with separation $\hbar\omega$,
with degeneracy formulae that are rather similar, albeit with the
CQO having higher degeneracies than those of the WQO. In fact the
analogy between them is somewhat closer if one compares the
infinite-dimensional CQO spectrum not with the finite-dimensional
spectrum associated with the WQO for any fixed, finite $p$, but with
the combination of all such WQO spectra for all non-negative
integer values of the order of the statistics  $p$.

Turning now to the angular momentum, it follows from~(\ref{MA})  and~(\ref{A+A-})  that
\bea && \hM_{\a1}\|p;.\t_{\a1},\t_{\a2},\t_{\a3}.\ra =i(\t_{\a2}-\t_{\a3})
\|p;.\t_{\a1},\ov{\t}_{\a2},\ov{\t}_{\a3},.\ra;
\label{Ma1}\\
&& \hM_{\a2}\|p;.\t_{\a1},\t_{\a2},\t_{\a3}.\ra =i(-1)^{\t_{\a2}}(\t_{\a3}-\t_{\a1})
\|p;.\ov{\t}_{\a1},\t_{\a2},\ov{\t}_{\a3}.\ra;\label{Ma2}\\
&& \hM_{\a3}\|p;.\t_{\a1},\t_{\a2},\t_{\a3}.\ra =i(\t_{\a1}-\t_{\a2})
\|p;.\ov{\t}_{\a1},\ov{\t}_{\a2},\t_{\a3}.\ra. \label{Ma3} \eea By exploiting these
results one then obtains \beq
   \hat{\M}_\a^2\|p;\T\ra = \d_\a\, 2\, \|p;\T\ra,
     \label{Ma^2}
\eeq
where
\beq
   \d_\a=\cases{0&if $\t_{\a1}=\t_{\a2}=\t_{\a3}$;\cr
               1&otherwise.\cr}
   \label{01}
\eeq
Thus the stationary states $\|p;\T\ra$ are eigenstates of the squares,
$\hat{\M}_\a^2$, of the {\it single particle} angular momentum operator $\hat{\M}_\a$,
with eigenvalues $0$ or $2$ for all $\a=1,2,\ldots,n$. Thus the WQO behaves like a
collection of spin zero and spin one particles.

However, the stationary states $\|p;\T\ra$ are not eigenstates of either $\hM_3$ or
$\hat{\M}^2$, the third component and the square, respectively, of the {\it total}
angular momentum operator $\hat{\M}$, as can be seen from the following: \bea &&
\hM_3\|p;\T\ra =i\sum_{\a =1}^n (\t_{\a 1}-\t_{\a 2}) \|p;.\ov{\t}_{\a 1},\ov{\t}_{\a
2},\t_{\a 3}.\ra.
\label{M3}\\
&& \hat{\M}^2\|p;\T\ra
=\sum_{\a=1}^n \d_\a\; 2\;\|p;\T\ra\nn \\
&& -2\sum_{1\leq\a<\b\leq n}  (\t_{\a1}\;-\t_{\a2})(\t_{\b1}\;-\t_{\b2})
\|p;.\ov{\t}_{\a1},\ov{\t}_{\a2},\t_{\a3},
\ldots,\ov{\t}_{\b1},\ov{\t}_{\b2},\t_{\b3}.\ra\nn \\
&& -2\sum_{1\leq\a<\b\leq n}  (-1)^{\t_{\a2}+\t_{\b2}}
(\t_{\a3}-\t_{\a1})(\t_{\b3}-\t_{\b1})
\|p;.\ov{\t}_{\a1},\t_{\a2},\ov{\t}_{\a3},
\ldots,\ov{\t}_{\b1},\t_{\b2},\ov{\t}_{\b3}.\ra\nn \\
&& -2\sum_{1\leq\a<\b\leq n}  (\t_{\a2}-\t_{\a3})(\t_{\b2}-\t_{\b3})
\|p;.\t_{\a1},\ov{\t}_{\a2},\ov{\t}_{\a3},
\ldots,\t_{\b1},\ov{\t}_{\b2},\ov{\t}_{\b3}.\ra. \label{M^2} \eea

For general values of the particle number $n$ it is not easy to
determine from these expressions all the total angular momentum
eigenstates, that is the simultaneous eigenvectors
of $\hM_3$ and $\hat{\M}^2$. However, to determine the
possible values of the total angular momentum, $M$, for the WQO
we can proceed in a different way by extending further our
restriction~(\ref{branch}) in accordance with the chain:
\beq
    sl(1|3n)\rightarrow gl(1)\oplus sl(3n)
\rightarrow gl(1) \oplus sl(3) \oplus sl(n)
\rightarrow gl(1) \oplus so(3) \oplus sl(n)
\rightarrow gl(1) \oplus so(3),
     \label{sl13n}
\eeq where it is the subalgebra $so(3)$ which is associated with the total angular
momentum of the system. The branching rule for $sl(3n)\rightarrow sl(3)\oplus sl(n)$
required in the second step and that for $sl(3)\rightarrow so(3)$ required in the third
step are both rather well known and have been implemented for example in
SCHUR\footnote{SCHUR, an interactive program for calculating properties of Lie groups
and symmetric functions, distributed by S. Christensen. E-mail: steve@scm.vnet.net;
http//scm.vnet.net/Christensen.html}. Since they involve coefficients for which there is
no known general formula, we content ourselves with giving the results explicitly just
for the two cases $n=1$ and $n=2$.

In the case of $sl(1|3)$ we find:
\bea
&  sl(1\vert 3)\ &\longrightarrow\ gl(1)\oplus so(3)\nn \\
&  V_{sl(1|3)}^p\ & \longrightarrow\
  \chi_{p\geq0}\ V_{gl(1)}^{3p} \otimes V_{so(3)}^0
  + \chi_{p\geq1}\ V_{gl(1)}^{3p-2} \otimes V_{so(3)}^1 \nn \\
 & &\qquad+ \chi_{p\geq2}\ V_{gl(1)}^{3p-4} \otimes V_{so(3)}^1
  + \chi_{p\geq3}\ V_{gl(1)}^{3p-6} \otimes V_{so(3)}^0, \label{sl13}
\eea
where $\chi_{p\geq x}$ is $1$ if $p\geq x$ and $0$ otherwise.
Each term of the form $V_{gl(1)}^{3p-2q}\otimes V_{so(3)}^M$
corresponds to a set of $2M+1$ states of energy
$E_q=\hbar\omega(3p-2q)/2$, as given by~(\ref{Eq})  with $n=1$, all
having total angular momentum $M$.

In this one particle, $n=1$, case it is easy to identify from~(\ref{Ma1})-(\ref{01})
with $\a=1$ all the angular momentum eigenstates, that is the simultaneous eigenvectors
of $\hat{\M}^2$ and $\hM_3$. They are the linear combinations of the stationary states
$\|p;\T\ra$ identified in Table~1.

\smallskip
\noindent{\bf Table~1. One-particle eigenstates of angular momentum}
$$\vbox{
\halign{\ $#$\qquad &$#$\qquad\hfil&$#$\qquad\hfil&\qquad$#$\hfil&\qquad$#$
\hfil
               &\qquad\hfil$#$&\hfil$#$\cr
\noalign{\hrule} \cr p\geq q&\|p;\T\ra&q&E_q&M&M_3\cr \cr \noalign{\hrule} \cr p\geq0&
\|p;0,0,0\ra&0&{{\hbar\omega}\over2} 3p&0&0\cr \cr
p\geq1&{1\over\sqrt2}(\|p;1,0,0\ra+i\|p;0,1,0\ra) &1&{{\hbar\omega}\over2}(3p-2)&1&1\cr
p\geq1&\|p;0,0,1\ra&1&{{\hbar\omega}\over2}(3p-2)&1&0\cr
p\geq1&{1\over\sqrt2}(\|p;1,0,0\ra-i\|p;0,1,0\ra) &1&{{\hbar\omega}\over2}(3p-2)&1&-1\cr
\cr p\geq2&{1\over\sqrt2}(\|p;1,0,1\ra+i\|p;0,1,1\ra)
&2&{{\hbar\omega}\over2}(3p-4)&1&1\cr
p\geq2&\|p;1,1,0\ra&2&{{\hbar\omega}\over2}(3p-4)&1&0\cr
p\geq2&{1\over\sqrt2}(\|p;1,0,1\ra-i\|p;0,1,1\ra) &2&{{\hbar\omega}\over2}(3p-4)&1&-1\cr
\cr p\geq3&\|p;1,1,1\ra&3&{{\hbar\omega}\over2}(3p-6)&0&0\cr \cr \noalign{\hrule} }}
$$

Thus in the atypical cases, $p=0$, $p=1$ and $p=2$ it is easy to see that
the dimensions of the corresponding irreducible representations of $sl(1|3)$
are $1$, $4$ and $7$, respectively, while for the typical cases $p\geq 3$ the
dimension is $8$, all in accordance with~(\ref{dim}).

In the two particle case, that is for $sl(1\vert 6)$ we find:
\bea
 & sl(1\vert 6) &\longrightarrow
  gl(1)\oplus so(3)\oplus sl(2)\nn \\
 &  V_{sl(1|6)}^p &\longrightarrow
  \chi_{p\geq0}\ V_{gl(1)}^{6p}\otimes
 \left(V_{so(3)}^0\otimes V_{sl(2)}^0\right)
  + \chi_{p\geq1}\ V_{gl(1)}^{6p-5}\otimes
\left(V_{so(3)}^1\otimes V_{sl(2)}^1\right) \nn\\
&& + \chi_{p\geq2}\ V_{gl(1)}^{6p-10}\otimes
\left(V_{so(3)}^2\otimes V_{sl(2)}^0+ V_{so(3)}^1\otimes V_{sl(2)}^2+
V_{so(3)}^0\otimes V_{sl(2)}^0\right)\nn\\
  &&
+ \chi_{p\geq3}\ V_{gl(1)}^{6p-15}\otimes
\left(V_{so(3)}^2\otimes V_{sl(2)}^1+ V_{so(3)}^1\otimes V_{sl(2)}^1+
V_{so(3)}^0\otimes V_{sl(2)}^3\right)\nn\\
  &&
+ \chi_{p\geq4}\ V_{gl(1)}^{6p-20}\otimes
\left(V_{so(3)}^2\otimes V_{sl(2)}^0+ V_{so(3)}^1\otimes V_{sl(2)}^2+
V_{so(3)}^0\otimes V_{sl(2)}^0\right)\label{sl16}\\
  &&
+ \chi_{p\geq5}\ V_{gl(1)}^{6p-25}\otimes
\left(V_{so(3)}^1\otimes V_{sl(2)}^1\right)
  + \chi_{p\geq6}\ V_{gl(1)}^{6p-30}\otimes
\left(V_{so(3)}^0\otimes V_{sl(2)}^0\right). \nn
\eea
Since the dimension of each irreducible representation
$V_{sl(2)}^s$ of $sl(2)$ is just $s+1$, it follows that:
\bea
  & sl(1\vert 6)\ &\longrightarrow\
  gl(1)\oplus so(3)\nn\\
  & V_{sl(1|6)}^p\ &\longrightarrow\
  \chi_{p\geq0}\ V_{gl(1)}^{6p}\otimes V_{so(3)}^0
  + \chi_{p\geq1}\ V_{gl(1)}^{6p-5}\otimes 2 V_{so(3)}^1 \nn\\
  &&\quad+ \chi_{p\geq2}\ V_{gl(1)}^{6p-10}\otimes
\left(V_{so(3)}^2 + 3 V_{so(3)}^1 + V_{so(3)}^0\right)\nn\\
  &&\quad+ \chi_{p\geq3}\ V_{gl(1)}^{6p-15}\otimes
\left(2 V_{so(3)}^2 + 2 V_{so(3)}^1 + 4 V_{so(3)}^0\right)\nn\\
  &&\quad+ \chi_{p\geq4}\ V_{gl(1)}^{6p-20}\otimes
\left(V_{so(3)}^2 + 3 V_{so(3)}^1 + V_{so(3)}^0\right)\nn\\
  &&\quad+ \chi_{p\geq5}\ V_{gl(1)}^{6p-25}\otimes 2 V_{so(3)}^1
  + \chi_{p\geq6}\ V_{gl(1)}^{6p-30}\otimes V_{so(3)}^0,\label{sl16p}
\eea
where now each term of the form $V_{gl(1)}^{6p-5q}\otimes k
V_{so(3)}^M$ corresponds to $k$ sets of $2M+1$ states of energy
$E_q=\hbar\omega(6p-5q)/5$, as given by~(\ref{Eq})  with $n=2$, all
having total angular momentum $M$.

Of course, for any particular value of $p<6$ not all of the above
terms will survive, as can be seen from the various factors
$\chi_{p\geq x}$. For example if $n=2$ and $p=3$ we obtain
\bea
 & sl(1\vert 6)\ &\longrightarrow\ gl(1)\oplus so(3)\nn\\
 & V_{sl(1|6)}^3\ &\longrightarrow\ V_{gl(1)}^{18}\otimes V_{so(3)}^0
  + V_{gl(1)}^{13}\otimes 2 V_{so(3)}^1 + V_{gl(1)}^{8}\otimes
\left(V_{so(3)}^2 + 3 V_{so(3)}^1 + V_{so(3)}^0\right)\nn\\
  && \qquad+ V_{gl(1)}^{3}\otimes
\left(2 V_{so(3)}^2 + 2 V_{so(3)}^1 + 4 V_{so(3)}^0\right).\label{sl163}
\eea

In the two particle, $n=2$, case it is not quite so easy to identify all the
eigenstates of both $\hat{\M}^2$ and $\hM_3$. In general they are now certain linear
combinations of the stationary states $\|p;\T\ra$. Rather than give all $64$ such
linear combinations, we content ourselves with specifying in Table~2 only those
eigenstates of $\hat{\M}^2$ with eigenvalues $M(M+1)$ for which $M_3=M$. The remaining
states may be obtained from these through the action of $\hM_-$, where \beq
   \hM_\pm=\hM_1\pm i\hM_2=\sum_{\a=1}^n \hM_{\a\pm}
\;\;\; \hbox{with}\ \hM_{\a\pm}=\hM_{\a1}\pm i\hM_{\a2}.
    \label{M+-}
\eeq

\vfill\eject

\noindent{\bf Table~2. Two-particle eigenstates of angular momentum having
the maximum value $M$ of $M_3$}
$$\vbox{
\halign{
      $#$\hfil 
&\quad$#$\hfil 
&\quad$#$\hfil 
&\quad$#$\hfil 
&\quad$#$\hfil 
&\quad$#$\hfil 
\cr \noalign{\hrule} \cr p\geq q &\hbox{Orthonormal angular momentum eigenstates} &q &\
E_q &\!\!\!M &\!\!\!M_3\cr &\hbox{with $M=M_3$ as linear combinations of $\|p;\T\ra$}\cr
\cr \noalign{\hrule} \cr p\geq0&\|p;000000\ra&0&{{\hbar\omega}\over5} 6p&0&0\cr \cr
p\geq1&{1\over\sqrt{2}}
(\|p;010000\ra-i\|p;100000\ra)&1&{{\hbar\omega}\over5}(6p-5)&1&1\cr
p\geq1&{1\over\sqrt{2}}
(\|p;000010\ra-i\|p;000100\ra)&1&{{\hbar\omega}\over5}(6p-5)&1&1\cr \cr p\geq2 &{1\over
2}(\|p;010010\ra-i\|p;100010\ra-i\|p;010100\ra-\|p;100100\ra)
&2&{{\hbar\omega}\over5}(6p-10)&2&2\cr p\geq2
&{1\over\sqrt{2}}(\|p;011000\ra-i\|p;101000\ra) &2&{{\hbar\omega}\over5}(6p-10)&1&1\cr
p\geq2 &{1\over\sqrt{2}}(\|p;000011\ra-i\|p;000101\ra)
&2&{{\hbar\omega}\over5}(6p-10)&1&1\cr p\geq2 &{1\over
2}\left(\|p;010001\ra-\|p;001010\ra+i\|p;001100\ra-i\|p;100001\ra\right)
&2&{{\hbar\omega}\over5}(6p-10)&1&1\cr p\geq2
&{1\over\sqrt3}\left(\|p;100100\ra+\|p;010010\ra+\|p;001001\ra \right)
&2&{{\hbar\omega}\over5}(6p-10)&0&0\cr \cr p\geq3 &{1\over
2}(\|p;010011\ra-\|p;100101\ra-i\|p;010101\ra-i\|p;100011\ra) &3
&{{\hbar\omega}\over5}(6p-15)&2&2\cr p\geq3 &{1\over
2}(\|p;011010\ra-\|p;101100\ra-i\|p;011100\ra-i\|p;101010\ra) &3
&{{\hbar\omega}\over5}(6p-15)&2&2\cr p\geq3 &{1\over
2}\left(\|p;001011\ra-\|p;100110\ra-i\|p;001101\ra-i\|p;010110\ra\right)
&3&{{\hbar\omega}\over5}(6p-15)&1&1\cr p\geq3 &{1\over
2}\left(\|p;011001\ra-\|p;110100\ra-i\|p;110010\ra-i\|p;101001\ra\right)
&3&{{\hbar\omega}\over5}(6p-15)&1&1\cr p\geq3 &\|p;000111\ra
&3&{{\hbar\omega}\over5}(6p-15)&0&0\cr p\geq3 &\|p;111000\ra
&3&{{\hbar\omega}\over5}(6p-15)&0&0\cr p\geq3
&{1\over\sqrt3}\left(\|p;100011\ra+\|p;001110\ra-\|p;010101\ra \right)
&3&{{\hbar\omega}\over5}(6p-15)&0&0\cr p\geq3
&{1\over\sqrt3}\left(\|p;011100\ra+\|p;110001\ra-\|p;101010\ra \right)
&3&{{\hbar\omega}\over5}(6p-15)&0&0\cr \cr p\geq4 &{1\over
2}(\|p;011011\ra-\|p;101101\ra-i\|p;011101\ra-i\|p;101011\ra)
&4&{{\hbar\omega}\over5}(6p-20)&2&2\cr p\geq4
&{1\over\sqrt{2}}(\|p;010111\ra-i\|p;100111\ra) &4&{{\hbar\omega}\over5}(6p-20)&1&1\cr
p\geq4 &{1\over\sqrt{2}}(\|p;111010\ra-i\|p;111010\ra)
&4&{{\hbar\omega}\over5}(6p-20)&1&1\cr p\geq4 &{1\over
2}\left(\|p;011110\ra-\|p;110011\ra-i\|p;101110\ra+i\|p;110101\ra\right)
&4&{{\hbar\omega}\over5}(6p-20)&1&1\cr p\geq4
&{1\over\sqrt3}\left(\|p;101101\ra+\|p;110110\ra+\|p;011011\ra \right)
&4&{{\hbar\omega}\over5}(6p-20)&0&0\cr \cr p\geq5
&{1\over\sqrt{2}}(\|p;011111\ra-i\|p;101111\ra) &5&{{\hbar\omega}\over5}(6p-25)&1&1\cr
p\geq5 &{1\over\sqrt{2}}(\|p;111011\ra-i\|p;111101\ra)
&5&{{\hbar\omega}\over5}(6p-25)&1&1\cr \cr p\geq6 &\|p;111111\ra
&6&{{\hbar\omega}\over5}(6p-30)&0&0\cr \cr \noalign{\hrule} }}
$$

\bigskip

In the case of the CQO, the results analogous to~(\ref{sl13})  and~(\ref{sl16p})
take the form
\bea
 & osp(1\vert 6)\ &\longrightarrow\ gl(1)\oplus so(3)\nn\\
 &  V_{osp(1|6)}^\varepsilon\ &\longrightarrow\
   V_{gl(1)}^{3/2} \otimes V_{so(3)}^0
  + V_{gl(1)}^{5/2} \otimes V_{so(3)}^1
  + V_{gl(1)}^{7/2} \otimes \left(V_{so(3)}^2 + V_{so(3)}^0\right)\nn\\
  &&\qquad+ V_{gl(1)}^{9/2} \otimes \left(V_{so(3)}^3+ V_{so(3)}^1\right)
+ \cdots \label{osp16}
\eea
where each term of the form $V_{gl(1)}^{(3+2q)/2}\otimes
V_{so(3)}^M$ corresponds to a set of $2M+1$ states of energy
$E_q=\hbar\omega(3+2q)/2$, as given by~(\ref{BEq})  with $n=1$, all
having total angular momentum $M$.

Similarly,
\bea
 & osp(1\vert 12) &\longrightarrow\ gl(1)\oplus so(3)\nn\\
 &  V_{osp(1|12)}^\varepsilon &\longrightarrow\
   V_{gl(1)}^{3} \otimes V_{so(3)}^0
  + V_{gl(1)}^{4} \otimes 2 V_{so(3)}^1
  + V_{gl(1)}^{5} \otimes
\left(3 V_{so(3)}^2 + V_{so(3)}^1 + 3 V_{so(3)}^0\right)\nn\\
  &&\qquad+ V_{gl(1)}^{6} \otimes
\left(4 V_{so(3)}^3+ 2 V_{so(3)}^2 + 6 V_{so(3)}^1\right) + \cdots
\label{osp112}
\eea
where now each term of the form $V_{gl(1)}^{3+q}\otimes k
V_{so(3)}^M$ corresponds to  $k$ sets of $2M+1$ states of energy
$E_q=\hbar\omega(3+q)$, as given by~(\ref{BEq})  with $n=2$, all having
total angular momentum $M$. It is notable that in the CQO case
the degeneracies are larger than in the case of the WQO, and of
course the CQO case is infinite-dimensional as compared with the
fixed $p$ finite-dimensional case of the WQO, illustrated for
example by~(\ref{sl163}).

\section{Physical properties - oscillator configurations}
\setcounter{equation}{0}

It is convenient to work not with the time dependent
position operators $\hR_{\a k}(t)$ themselves, but with their
dimensionless version defined by
\beq
\hr_{\a k}(t)={\sqrt {(3n-1)m\omega\over \hbar}}\hR_{\a k}(t)
 = A_{\a k}^+\, e^{-i\omega t} + A_{\a k}^-\, e^{i\omega t},
\label{rt}
\eeq
for $k=1,2,3$ and $\a=1,2,\ldots,n$.
It then follows from~(\ref{sl})  that the squares of these
operators are time independent and given by
\beq
  \hr_{\a k}^2=\{A_{\a k}^+,A_{\a k}^-\}.
\label{r^2}
\eeq
The first part of~(\ref{A+A-})  then implies that
\beq
\hr_{\a k}^2 \|p;\T\ra = r_{\a k}^2 \|p;\T\ra,
\label{r^2b}
\eeq
with
\beq
  r_{\a k}^2=p-q+\t_{\a k}.
\label{r^2ev}
\eeq
Since
\beq
   [\hr_{\a i}^2,\hr_{\b j}^2]=0 \quad\hbox{for all $\a,\b=1,2,\ldots,n$
   and $i,j=1,2,3$},
   \label{ri^2rj^2}
\eeq
we are led to the following:

\begin{conc}
If the system is in one of the $\T$-basis states
$\|p;\T\ra$, then measurements of the coordinates $r_{\a k}$
of the $\a$th particle can yield only the values
\beq
 r_{\a k}=\pm \sqrt{p-q+\t_{\a k}}
 \quad\hbox{for $k=1,2,3$}.
  \label{rev}
\eeq
These values define the position of $8$ nests
on a sphere of radius
\beq
\rho_\a=\sqrt{3p-3q+q_\a}\quad\hbox{where $q_\a=\sum_{k=1}^3\t_{\a k}$}.
 \label{rho}
\eeq
\end{conc}

This result is completely analagous to that found in the single particle case, (cf.
{\tt I} Conclusion 2).

The basis states $\|p;\T\ra$ are not eigenstates of $\hr_{\a k}(t)$.
In fact
\beq
   \hr_{\a k}(t) \|p;\ldots,\t_{\a k},\ldots\ra  =(-1)^{\psi_{\a k}}
\left( e^{-i\omega t}\ov{\t}_{\a k}+e^{i\omega t} \t_{\a k} \right)
 \sqrt{p-q+\t_{\a k}} \|p;\ldots,\ov{\t}_{\a k},\ldots\ra
 \label{rbasis}
\eeq
However, it is not difficult to identify in $W(n,p)$ the eigenvectors of
$\hr_{\a k}(t)$ for any $\a =1,\ldots ,n$ and  $k=1,2,3$.
These are given by:
\beq
v_{\a k}(\T) = {1\over \sqrt{2}}\big( \|p;\T_{\t_{\a k}=0}\ra +
(-1)^{\psi_{\a k}+\t_{\a k}} e^{-i\omega t} \|p;\T_{\t_{\a k=1}}\ra\big),
\label{v}
\eeq
where $\T_{\t_{\a k}=0}$ stands for the $\T$-value specified by the
left hand side of~(\ref{v}) in which $\t_{\a k}$ is replaced by~$0$
(and similarly for $\T_{\t_{\a k}=1}$). The vectors $v_{\a k}(\T)$
constitute an orthonormal basis of eigenvectors of $\hr_{\a k}(t)$ in
$W(n,p)$. It is found that
\beq
\hr_{\a k}(t)v_{\a k}(\T)
= (-1)^{\t_{\a k}} \sqrt{p-q+\t_{\a k}} v_{\a k}(\T),
\label{rv}
\eeq
confirming the fact that the eigenvalues of $\hr_{\a k}(t)$
are given by $\pm\sqrt{p-q+\t_{\a k}}$.
The inverse relations of~(\ref{v}) are easy to write down. They take the
form:
\beq
\|p;\T\ra = {1\over\sqrt{2}} (-1)^{\psi_{\a k}
\t_{\a k}} e^{i\omega t \t_{\a k}}
\big( v_{\a k}(\T_{\t_{\a k}=0})
+ (-1)^{\t_{\a k}} v_{\a k}(\T_{\t_{\a k}=1})\big).
\label{inverse}
\eeq

As pointed out in {\tt I}, the interpretation of geometric results
for the WQO must be undertaken
carefully since the underlying geometry is non-commutative.
This can be seen by noting that for all $(\a i)<(\b j)$
with $\a,\b=1,2,\ldots,n$ and $i,j=1,2,3$
\bea
 &&[\hr_{\a i}(t),\hr_{\b j}(t)]
\|p;\ldots,\t_{a i},\ldots,\t_{\b j},\ldots\ra \nn\\
 &&= (-1)^{\psi_{\b j}-\psi_{\a i}} \biggl(2
  e^{i2\omega t} {\t}_{\a i}{\t}_{\b j} \sqrt{(p-q+1)(p-q+2)}
 +({\t}_{\a i}-{\t}_{\b j})^2(2p-2q+1) \nn\\
&&\quad + 2e^{-i2\omega t} \ov{\t}_{\a i}\ov{\t}_{\b j}\sqrt{(p-q-1)(p-q)}\biggr)
\|p;\ldots,\ov{\t}_{a i},\ldots,\ov{\t}_{\b j},\ldots\ra. \label{rairbj} \eea The right
hand side of this expression is non-zero for all $p\geq q+2$. This implies in
particular in the $\a=\b$ case that measurements of the $i$th and $j$th coordinates of
the $\a$th particle do not, in general, commute. Thus the position of the $\a$th
particle may not be specified precisely. The most that can be said is that for each
$\T$ with $p$ sufficiently large for the representation to be typical we can associate
with $\|p;\T\ra$ eight nests whose coordinates serve to specify the possible outcomes of
measurements of $r_{\a k}$ for $k=1,2,3$.

It is of course possible to take measurements not of the
coordinates $r_{\a k}(t)$ with respect to the original frame of
reference, but of coordinates $s_{\a k}(t)$ associated with
some alternative frame of reference whose orientation
with respect to the first may be specified by means, for
example, of certain Euler angles. For the sake of simplicity
to illustrate the issues involved we consider an orientation
obtained by rotating the frame of reference through an angle
$\phi$ about the third axis. The relevant position operators then take
the form:
\bea
&\hs_{\a1}(t)&=\cos\phi\,\hr_{\a1}(t)+\sin\phi\,\hr_{\a2}(t);\nn\\
& \hs_{\a2}(t)&=-\sin\phi\,\hr_{\a1}(t)+\cos\phi\,\hr_{\a2}(t);\label{sai}\\
&\hs_{\a3}(t)&=\hr_{\a3}(t)\nn.
\eea

Once again the squares of these operators mutually commute, they are
time independent and they commute with the Hamiltonian. They are given
by
\bea
&& \hs_{\a 1}^2=\cos^2\phi\,\{A_{\a 1}^+,A_{\a 1}^-\}
+\cos\phi\,\sin\phi\,
\left(\{A_{\a 1}^+,A_{\a 2}^-\}+\{A_{\a 2}^+,A_{\a 1}^-\}\right)
+\sin^2\phi\,\{A_{\a 2}^+,A_{\a 2}^-\};\nn\\
&&\label{sa1^2}\\
&& \hs_{\a 2}^2=\sin^2\phi\,
\{A_{\a 1}^+,A_{\a 1}^-\}
-\cos\phi\,\sin\phi\,
\left(\{A_{\a 1}^+,A_{\a 2}^-\}+\{A_{\a 2}^+,A_{\a 1}^-\}\right)
+\cos^2\phi\,\{A_{\a 2}^+,A_{\a 2}^-\};\nn\\
&&\label{sa2^2}\\
&&\hs_{\a 3}^2=\{A_{\a 3}^+,A_{\a 3}^-\}.
\label{sa3^2}
\eea

Their action on the stationary states is such that
\bea
&&\hs_{\a1}^2\|p;.\t_{\a1}\t_{\a2}\t_{\a3}.\ra =\left(p-q+\t_{\a1}\cos^2\phi
+\t_{\a2}\sin^2\phi\right)
\|p;.\t_{\a1}\t_{\a2}\t_{\a3}.\ra \nn\\
&&\qquad\qquad +(\t_{\a1}-\t_{\a2})^2\cos\phi\,\sin\phi\,
\|p;.\ov{\t}_{\a1}\ov{\t}_{\a2}\t_{\a3}. \ra;
\label{sa1^2b}\\
&& \hs_{\a2}^2\|p;.\t_{\a1}\t_{\a2}\t_{\a3}.\ra
=\left(p-q+\t_{\a1}\sin^2\phi\,+\t_{\a2}\cos^2\phi\,\right)
\|p;.\t_{\a1}\t_{\a2}\t_{\a3}.\ra \nn\\
&&\qquad\qquad -(\t_{\a1}-\t_{\a2})^2\cos\phi\,\sin\phi\,
\|p;.\ov{\t}_{\a1}\ov{\t}_{\a2}\t_{\a3}. \ra;
\label{sa2^2b} \\
&&\hs_{\a3}^2\|p;.\t_{\a1}\t_{\a2}\t_{\a3}.\ra =\left(p-q+\t_{\a3}\right)
\|p;.\t_{\a1}\t_{\a2}\t_{\a3}.\ra, \label{sa3^2b} \eea where $\ov{\t}_{\a i}=1-\t_{\a
i}$ for $i=1,2,3$.

Clearly the states $\|p;\T\ra$ are not eigenstates of $\hs_{\a k}^2$. However, it is not
difficult to identify the common eigenstates of these mutually commuting operators.
They are given by \ \beq \vbox{ \halign{$#$\hfill&\quad#\hfill\cr
 {\phantom{+\cos\phi\,}}\|p;\cdot\t_{\a1}\t_{\a2}\t_{\a3}\cdot\ra
 &for $\t_{\a1}=\t_{\a2}=0$ and $\t_{\a1}=\t_{\a2}=1$;\cr
 {\ph+}\cos\phi\,\|p;\cdot\t_{\a1}\t_{\a2}\t_{\a3}\cdot\ra
 +\sin\phi\,\|p;\cdot\ov{\t}_{\a1}\ov{\t}_{\a2}\t_{\a3}\cdot\ra
 &for $\t_{\a1}=1,\t_{\a2}=0$;\cr
 -\sin\phi\,\|p;\cdot\t_{\a1}\t_{\a2}\t_{\a3}\cdot\ra
 +\cos\phi\,\|p;\cdot\ov{\t}_{\a1}\ov{\t}_{\a2}\t_{\a3}\cdot\ra
 &for $\t_{\a1}=1,\t_{\a2}=0$.\cr\cr}}\label{es}
\eeq
The corresponding eigenvalues $s_{\a k}^2$
are as shown in Table~3.
\bigskip

\noindent{\bf Table~3. Eigenvalues and eigenvectors
of $\hs_{\a k}^2$}
$$\vbox{
\halign{
     $#$\ \hfill&
\quad$#$\ \hfill&
\quad$#$\ \hfill&
\quad$#$\ \hfill&
\quad$#$\hfill\cr
\cr
\noalign{\hrule}
\cr
\hbox{Eigenvectors expressed as linear}
&s_{\a1}^2&s_{\a2}^2&s_{\a3}^2&{\bf{s_\a}}^2\cr
\hbox{combinations of
$\|p;\cdot\t_{\a1}\t_{\a2}\t_{\a3}\cdot\ra$}
&\cr
\noalign{\hrule}
\cr
\|p;\cdot000\cdot\ra
&p-q  &p-q  &p-q  &3p-3q\cr
{\ph+}\cos\phi\,\|p;\cdot100\cdot\ra+\sin\phi\,\|p;\cdot010\cdot\ra
&p-q+1  &p-q&p-q&3p-3q+1\cr
-\sin\phi\,\|p;\cdot100\cdot\ra+\cos\phi\,\|p;\cdot010\cdot\ra
&p-q&p-q+1  &p-q&3p-3q+1\cr
\|p;\cdot001\cdot\ra
&p-q&p-q&p-q+1  &3p-3q+1\cr
\|p;\cdot110\cdot\ra
&p-q+1&p-q+1&p-q&3p-3q+2\cr
{\ph+}\cos\phi\,\|p;\cdot101\cdot\ra+\sin\phi\,\|p;\cdot011\cdot\ra
&p-q+1&p-q&p-q+1&3p-3q+2\cr
-\sin\phi\,\|p;\cdot101\cdot\ra+\cos\phi\,\|p;\cdot011\cdot\ra
&p-q&p-q+1&p-q+1&3p-3q+2\cr
\|p;\cdot111\cdot\ra
&p-q+1&p-q+1&p-q+1&3p-3q+3\cr
\noalign{\hrule}
\cr}}
$$

Notice that we have, as required,
\beq
   {\bf{s}}_{\a}^2=\sum_{k=1}^3 \s_{\a k}^2=3p-3q+\t_{\a 1}+\t_{\a
   2}+\t_{\a 3}=\rho_{\a}^2.
  \label{sa^2}
\eeq

The eigenvalues $s_{\a k}$ of $\hs_{\a k}(t)$ for $k=1,2,3$ are
just $\pm$ the square root of those tabulated for $\hs_{\a k}^2$.
These results indicate that the sites corresponding to possible values
of measurements of the coordinates $\s_{\a k}$ are again
nests on a sphere of radius $\rho_{\a}$, but the nests define a
rectangular parallelepiped obtain by rotating the original one
about the third axis through an angle $\phi$.

It is particularly striking that the measured values of $s_{\a k}$ are of the form
\break $\pm\sqrt{p-q+\t}$ with $\t\in\{0,1\}$, just as for $r_{\a k}$. They are not the
values one might have expected by looking at the nests defined with respect to
measurements of $r_{\a k}$. This is especially clear in the case of the common
eigenstate $\|p;.000.\ra$ of all the operators $\hr_{\a k}^2$ and $\hs_{\a k}^2$ for
$k=1,2,3$. The eigenvalues of $\hr_{\a k}$ and $\hs_{\a k}$  are all $\pm\sqrt{p-q}$.
Thus for example in the case $\phi =\pi/4$ the nests defined with respect to
measurements of $r_{\a k}$ for $k=1,2,3$ have coordinates $s_{\a
k}\in\{0,\pm\sqrt{2(p-q)}\}$ for $k=1$ and $2$ and all $\a$. These are not coordinates
of the nests defined with respect to measurements of $s_{\a k}$ for $k=1,2,3$. The
explanation for this lies in the fact that the particles themselves may not be
localised, since measurements of their coordinates do not mutually commute. It is the
choice of coordinate to be measured that leads to the observed value corresponding to
the associated eigenvalue. This is analagous to the ordinary quantum mechanical
measurement of angular momentum, whereby a state of total angular momentum $J$ is such
that measurements of the third component of angular momentum gives rise to a discrete
set of possible values $J_3=J,J-1,\ldots,-J$ regardless of the orientation of the third
axis. For example, measurements on a particle of spin ${1\over2}$ yield values for the
projection of the spin in any given direction of only $\pm{1\over2}$. The result is
never $0$ as might have been expected in a direction perpendicular to the direction in
which it is observed to have spin projection ${1\over2}$, nor ${1\over2}\cos\phi $ for
any rotation of the axes through an angle $\phi$.

These observations regarding measurements of the coordinates $s_{\a k}$
may be generalized to the case of coordinates obtained from $r_{\a k}$
by means of any orthogonal transformation in the underlying 3D space.
For each $g^\a\in O(3)$ let
$g^\a:\hbr_{\a}(t)\mapsto\hbs_{\a}(t)=\hbr_{\a}(t) g^\a$,
so that for $g^\a=(g^\a_{ij})_{1\leq i,j\leq 3}$ we have
\beq
  \hs_{\a k}(t)=\sum_{i=1}^3 \hr_{\a i}(t) g^\a_{ik},
\label{sak}
\eeq
with
\beq
  \sum_{k=1}^3 g^\a_{ik} g^\a_{jk} = \delta_{ij}.
 \label{g}
\eeq
Just as in the case of~(\ref{sai}), the squares of the operators~(\ref{sak})
mutually commute, they are time independent and they commute with
the Hamiltonian. Their action on the stationary states
$\|p;\T\ra=\|p;.\t_{\a1}\t_{\a2}\t_{\a3}.\ra$ is such that
\bea
&&\hs_{\a k}^2\|p;.\t_{\a1}\t_{\a2}\t_{\a3}.\ra\nn\\
&&\quad =\left(p-q+(g^\a_{1k})^2\t_{\a 1}+(g^\a_{2k})^2\t_{\a 2}
+(g^\a_{3k})^2\t_{\a 3}\right)
\|p;.\t_{\a1}\t_{\a2}\t_{\a3}.\ra \nn\\
&&\qquad\qquad +g^\a_{1k}g^\a_{2k}\,(\t_{\a1}-\t_{\a2})^2
\|p;.\ov{\t}_{\a1}\ov{\t}_{\a2}\t_{\a3}. \ra\nn\\
&&\qquad\qquad +g^\a_{1k}g^\a_{3k}(-1)^{\t_{\a 2}}\,(\t_{\a1}-\t_{\a3})^2
\|p;.\ov{\t}_{\a1}{\t}_{\a2}\ov{\t}_{\a3}. \ra\nn\\
&&\qquad\qquad +g^\a_{2k}g^\a_{3k}\,(\t_{\a2}-\t_{\a3})^2
\|p;.{\t}_{\a1}\ov{\t}_{\a2}\ov{\t}_{\a3}. \ra. \label{sak^2} \eea As expected the
states $\|p;\T\ra$ are not eigenstates of $\hs_{\a k}^2$. The common eigenstates of
these mutually commuting operators are given in Table~4, along with the eigenvalues
\bigskip

\noindent{\bf Table~4. Eigenvalues and eigenvectors
of $\hs_{\a k}^2$}
$$\vbox{
\halign{
     $#$\ \hfill&
\quad$#$\ \hfill&
\quad$#$\ \hfill&
\quad$#$\ \hfill&
\quad$#$\hfill\cr
\cr
\noalign{\hrule}
\cr
&\hbox{Eigenvectors expressed as linear}
&s_{\a k}^2&{\bf{s}_\a}^2\cr
&\hbox{combinations of
$\|p;\cdot\t_{\a1}\t_{\a2}\t_{\a3}\cdot\ra$}
&k=1,2,3&\cr
\noalign{\hrule}
\cr
&\|p;\cdot000\cdot\ra
&p-q   &3p-3q\cr
j=1,2,3&g^\a_{1j}\,\|p;\cdot100\cdot\ra+g^\a_{2j}\,\|p;\cdot010\cdot\ra
+g^\a_{3j}\,\|p;\cdot001\cdot\ra
&p-q+\delta_{jk}&3p-3q+1\cr
j=1,2,3&g^\a_{1j}\,\|p;\cdot011\cdot\ra-g^\a_{2j}\,\|p;\cdot101\cdot\ra
+g^\a_{3j}\,\|p;\cdot110\cdot\ra
&p-q+1-\delta_{jk}&3p-3q+2\cr
&\|p;\cdot111\cdot\ra
&p-q+1&3p-3q+3\cr
\noalign{\hrule}
\cr}}
$$

As can be seen, the eigenvalues $s_{\a k}^2$ take the values
$p-q+\t$ with $\t\in\{0,1\}$. It follows that
\beq
     s_{\a k}=\pm\sqrt{p-q+\t}
\quad\hbox{with $\t\in\{0,1\}$},
 \label{sakev}
\eeq
so that we have the usual set of eight nests for the $\a$th particle
with respect this time to measurements of the coordinates
$s_{\a k}$ for $k=1,2,3$ obtained from $r_{\a k}$ by means of
the orthogonal transformation $g^\a$.

It is particularly noteworthy that the states $\|p;\cdot000\cdot\ra$ and
$\|p;\cdot111\cdot\ra$ are eigenstates of $\hat{s}_{\a k}^2$ for all $g^\a\in O(3)$,
that is to say for all choices of coordinates $s_{\a k}$. It follows that for these two
particular states, $\|p;\cdot000\cdot\ra$ and $\|p;\cdot111\cdot\ra$, the two sets of
eight nests with coordinates $\pm\sqrt{p-q}$ and $\pm\sqrt{p-q+1}$ can appear anywhere
on the spheres of radii $\pm\sqrt{3p-3q}$ and $\pm\sqrt{3p-3q+3}$, respectively. The
other states $\|p;\cdot\t_{\a1}\t_{\a2}\t_{\a3}\cdot\ra$ with $\t_{\a i}\neq\t_{\a  j}$
for some $i\neq j$, are not eigenstates of $\hat{s}_{\a k}^2$ for all $g^\a\in O(3)$,
as can be seen from Table~4 in which the $g^\a$ dependent eigenstates are specified.
This time it is these eigenstates which define two sets of eight nests on the spheres
of radii $\pm\sqrt{3p-3q+1}$ and $\pm\sqrt{3p-3q+2}$, oriented in accordance with the
specification of $s_{\a k}$ that is determined by $g^\a$. The fact, previously noted in
{\tt I}, that the geometry is non-commutative and the position of a particle is not well
defined, coupled with the existence of arbitrarily oriented sets of nests, makes the
interpretation of measurements of the position of particle $\a$ somewhat difficult.

In these circumstances, we must expect some difficulties
over the interpretation of the measurement of the
distance between two particles $\a$ and $\b$. We cannot
after all simultaneously specify their positions. However,
in line with classical notions of distance,
we are free to call $\hd_{\a\b}^2(t)$ the square distance
operator for particles $\a$ and $\b$, where
\beq
 \hd_{\a\b}^2(t)=\sum_{i=1}^3 \left(\hr_{\a i}(t)-\hr_{\b i}(t)\right)^2,
 \label{dab^2}
\eeq
and to examine its properties, including its spectrum of eigenvalues
in the space $W(n,p)$.

In the case $n=2$, with $\a=1$ and $\b=2$ we find that $\hd_{12}^2(t)$ has the
eigenvalues listed below along with their multiplicities specified by means of
subscripts:

\beq
\begin{array}{l|l|l}
 (6p)_1  & (6p-12)_4 & (6p-22)_3 \\[2mm]
 (6p-4)_3  & (6p-14)_9 &   (6p-24)_3 \\[2mm]
 (6p-6)_3  & (6p-16)_9 &  (6p-26)_3 \\[2mm]
 (6p-8)_3  & (6p-18)_4 &  (6p-30)_1 \\[2mm]
 (6p-10)_9  &   (6p-20)_9 &
\end{array} \label{mult}
\eeq

What is remarkable about these eigenvalues is that they do not coincide with the values
one might have naively expected, namely the squares of the distances between the
positions of the nests defined by $r_{1i}=\pm\sqrt{p-q+\t_{1i}}$ and
$r_{2j}=\pm\sqrt{p-q+\t_{2j}}$. For example, in the state $\|p;\T\ra=\|p;0,0,0,0,0,0\ra$
we have $r_{1i}=\pm\sqrt{p}$ and $r_{2i}=\pm\sqrt{p}$ for $i=1,2,3$. It follows that
\beq
   \sum_{i=1}^3 \left(r_{1i}-r_{2i}\right)^2
    \in\{0,4p,8p,12p\}.
\label{r1i-r2i} \eeq
Thus the spectrum of eigenvalues of $\hd_{12}^2(t)$ given
in~(\ref{mult}) does not contain for general $p$ the values of the squares of the
distances between nests as given in~(\ref{r1i-r2i}) for the
 particularly simple state $\|p;0,0,0,0,0,0\ra$. This
apparent contradiction leads one to ask if the results we have obtained for a
multi-particle Wigner quantum oscillator can possibly be self-consistent. The answer is
``Yes''. The explanation of the apparent disagreement stems from the non-commutativity
of measurements of the coordinates. Even though the state $\|p;0,0,0,0,0,0\ra$ is a
simultaneous eigenstate of $\hr_{1i}^2(t)$, $\hr_{2j}^2(t)$ and $\hd_{12}^2(t)$, the
positions of the nests are defined by the eigenvalues of $\hr_{1i}(t)$ and
$\hr_{2j}(t)$. These operators do not commute with $\hd_{12}^2(t)$. Morevover,
$\|p;0,0,0,0,0,0\ra$ is not one of their eigenstates. Hence it is not surprising that
the eigenvalues~(\ref{mult}) of $\hd_{12}^2(t)$ do not include the values given
in~(\ref{r1i-r2i}) for the squares of the distances between the nests.

Returning to the general case, in the space $W(n,p)$ what does remain well defined with
respect to measurements of the positions of particles $\a$ and $\b$ in the state
$\|p;\T\ra$ are the radii $\rho_\a$ and $\rho_\b$ of the spheres on which they are
located, namely \beq
  \rho_\a=\sqrt{3p-3q+q_\a}
  \quad\hbox{and}\quad
  \rho_\b=\sqrt{3p-3q+q_\b}.
\label{rhoab} \eeq In such a state, $\|p;\T\ra$, the expectation value of the square
distance operator for particles $\a$ and $\b$ is given by \beq
   \ov{d^2}_{\a\b}(t)=\la p;\T|\ \hd_{\a\b}^2(t)\ \|p;\T\ra,
\label{expv}
\eeq
where we can assume $\a<\b$ without loss of generality.

As in conventional quantum mechanical models and our postulate (P2) in
{\tt I}, we would
expect to interpret this as the average value of the square
of the distance between the particles $\a$ and $\b$.
To calculate this quantity is convenient to let
\beq
    \hd_{\a\b k}(t)= \hr_{\a k}(t)-\hr_{\b k}(t)
    \quad\hbox{for $k=1,2,3$ and $1\leq \a<\b\leq n$}.
  \label{dabk}
\eeq
Using~(\ref{rt})  the squares of these operators are given by
\bea
  &  \hd_{\a\b k}^2(t)
&= \hr_{\a k}^2-\{\hr_{\a k},\hr_{\b k}\}+\hr_{\b k}^2 \nn\\
&&= \{A_{\a k}^+,A_{\a k}^-\}-\{A_{\a k}^+,A_{\b k}^-\}
-\{A_{\b k}^+,A_{\a k}^-\}+\{A_{\b k}^+,A_{\b k}^-\},\label{dabk^2}
\eea
where use has been made of the last part of~(\ref{sl}).
Amongst other things this serves to eliminate
the time dependence from $\hd_{\a\b k}^2(t)$.
It then follows from~(\ref{A+A-})  that
\bea
 &&\hd_{\a\b k}^2 \|p;\ldots,\t_{\a k},\ldots,\t_{\b k},\ldots\ra
  = \left(2p-2q+\t_{\a k}+\t_{\b k}\right)
\|p;\ldots,\t_{\a k},\ldots,\t_{\b k},\ldots\ra\nn\\
 &&\qquad\qquad  + (-1)^{\psi_{\b k}-\psi_{\a k}} (\t_{\a k}-\t_{\b k})
\|p;\ldots,\ov{\t}_{\a k},\ldots,\ov{\t}_{\b k},\ldots\ra.\label{dabk^2b}
\eea
Hense
\bea
  & \ov{d^2}_{\a\b}&=\la p;\T|\ \hd_{\a\b}^2\ \|p;\T\ra
   =\sum_{k=1}^3 \la p;\T|\ \hd_{\a\b k}^2\ \|p;\T\ra\nn\\
  &&=\sum_{k=1}^3 (2p-2q+\t_{\a k}+\t_{\b k})
   =6p-6q+q_\a+q_\b=\rho_\a^2+\rho_\b^2,\label{expv1}
\eea
in the notation of~(\ref{rhoab}).

For consistency of interpretation we would then expect this to coincide with
$d_{\a\b}^2$, the average of the square of the distance between the nests available to
particles $\a$ and $\b$ in each of the states $\|p;\T\ra$. This average will depend of
course on the probabilities of occupying each nest.

Now suppose that the $n$ particle system is in one of the $\T$-basis states
$\|p;\T\ra$. Then the $\g$-th particle ($\g =1,\ldots n$) could have the following
coordinates: \beq r_{\g i}=\pm \sqrt{p-q+\t_{\g i}}. \label{rgammai} \eeq

Let $s=(\pm\pm\pm)$ be a sequence of signs specifying the sites of the $8$ possible
nests of the $\g$-th particle associated with a particular state $\|p;\T\ra$ by
signifying the signs of the corresponding coordinates $(r_{\g 1},r_{\g 2},r_{\g 3})$.
Let ${\cal{P}}_\g (s)$ be the probability of finding the $\g $-th particle in the nest
specified by $s$. Then from~\cite{KPSJ} we have \bea
&&  {\cal{P}}_\g(+++)+{\cal{P}}_\g(++-)+{\cal{P}}_\g(+-+)+{\cal{P}}_\g(+--)={1\over2}, \nn\\
&&  {\cal{P}}_\g(-++)+{\cal{P}}_\g(-+-)+{\cal{P}}_\g(--+)+{\cal{P}}_\g(---)={1\over2}, \nn\\
&&  {\cal{P}}_\g(+++)+{\cal{P}}_\g(++-)+{\cal{P}}_\g(-++)+{\cal{P}}_\g(-+-)={1\over2},
\label{prob}\\
&&  {\cal{P}}_\g(+-+)+{\cal{P}}_\g(+--)+{\cal{P}}_\g(--+)+{\cal{P}}_\g(---)={1\over2}, \nn\\
&&  {\cal{P}}_\g(+++)+{\cal{P}}_\g(+-+)+{\cal{P}}_\g(-++)+{\cal{P}}_\g(--+)={1\over2}, \nn\\
&&  {\cal{P}}_\g(++-)+{\cal{P}}_\g(+--)+{\cal{P}}_\g(-+-)+{\cal{P}}_\g(---)={1\over2}. \nn
\eea

The average square distance of the particle $\a$, occupying the nest at $(r_{\a
1},r_{\a 2},r_{\a 3})$ with probability ${\cal P}_\a(+++)$, from the particle $\b$,
occupying the 8 nests at \break
$(r_{\b 1\pm},r_{\b 2\pm},r_{\b 3\pm})$, with
probability ${\cal P}_\b(\pm\pm\pm)$, where $r_{\a k\pm}=\pm\sqrt{p-q-\t_{\a k}}$ and
$r_{\b k\pm}=\pm\sqrt{p-q-\t_{\b k}}$ for $k=1,2,3$, is then given by \bea
&&{\cal{P}}_\a(+++){\cal{P}}_\b(+++)\big( (r_{\a 1}-r_{\b 1})^2+
(r_{\a 2}-r_{\b 2})^2+(r_{\a 3}-r_{\b 3})^2\big)+ \nn\\
&&{\cal{P}}_\a(+++){\cal{P}}_\b(++-)\big( (r_{\a 1}-r_{\b 1})^2+
(r_{\a 2}-r_{\b 2})^2+(r_{\a 3}+r_{\b 3})^2\big)+ \nn\\
&&{\cal{P}}_\a(+++){\cal{P}}_\b(+-+)\big( (r_{\a 1}-r_{\b 1})^2+
(r_{\a 2}+r_{\b 2})^2+(r_{\a 3}-r_{\b 3})^2\big)+ \nn\\
&&{\cal{P}}_\a(+++){\cal{P}}_\b(+--)\big( (r_{\a 1}-r_{\b 1})^2+
(r_{\a 2}+r_{\b 2})^2+(r_{\a 3}+r_{\b 3})^2\big)+ \nn\\
&&{\cal{P}}_\a(+++){\cal{P}}_\b(-++)\big( (r_{\a 1}+r_{\b 1})^2+
(r_{\a 2}-r_{\b 2})^2+(r_{\a 3}-r_{\b 3})^2\big)+ \nn\\
&&{\cal{P}}_\a(+++){\cal{P}}_\b(-+-)\big( (r_{\a 1}+r_{\b 1})^2+
(r_{\a 2}-r_{\b 2})^2+(r_{\a 3}+r_{\b 3})^2\big)+ \nn\\
&&{\cal{P}}_\a(+++){\cal{P}}_\b(--+)\big( (r_{\a 1}+r_{\b 1})^2+
(r_{\a 2}+r_{\b 2})^2+(r_{\a 3}-r_{\b 3})^2\big)+ \nn\\
&&{\cal{P}}_\a(+++){\cal{P}}_\b(---)\big( (r_{\a 1}+r_{\b 1})^2+
(r_{\a 2}+r_{\b 2})^2+(r_{\a 3}+r_{\b 3})^2\big)\nn\\
&&={1\over 2}{\cal{P}}_\a(+++)\big( (r_{\a 1}-r_{\b 1})^2+
(r_{\a 2}-r_{\b 2})^2+(r_{\a 3}-r_{\b 3})^2\big) \nn\\
&&+{1\over 2}{\cal{P}}_\a(+++)\big( (r_{\a 1}+r_{\b 1})^2+
(r_{\a 2}+r_{\b 2})^2+(r_{\a 3}+r_{\b 3})^2\big)\nn\\
&&={\cal{P}}_\a(+++)\sum_{i=1}^3(r_{\a i}^2+r_{\b i}^2)=
{\cal{P}}_\a(+++)(\rho_\a^2+\rho_\b^2).\label{prob1}
\eea
In the crucial first step all the various parts of~(\ref{prob})  have been used
with $\g$ set equal to $\b$.

Treating the other seven nests for
the particle $\a$ in a similar way one can compute the average square
distance of particle $\a$, occupying the
8 nests at $(r_{\a 1\pm},r_{\a 2\pm},r_{\a 3\pm})$ with probabilities
${\cal P}_\a(\pm\pm\pm)$,
from particle $\b$, occupying the
8 nests at $(r_{\b 1\pm},r_{\b 2\pm},r_{\b 3\pm})$ with probabilities
${\cal P}_\b(\pm\pm\pm)$.
We arrive at the result
\bea
&&\bigg( {\cal{P}}_\a(+++)+{\cal{P}}_\a(++-)+{\cal{P}}_\a(+-+)+{\cal{P}}_\a(+--)+ \nn\\
&&\qquad
{\cal{P}}_\a(-++)+{\cal{P}}_\a(-+-)+{\cal{P}}_\a(--+)+{\cal{P}}_\a(---)\bigg)\big
(\rho_\a^2+\rho_\b^2\big )\nn\\
&&=\rho_\a^2+\rho_\b^2=6p-6q+q_\a +q_\b,\label{prob2}
\eea
in perfect agreement with~(\ref{expv1}).
It is notable, as can be seen from~(\ref{prob1}), that
even if the particle $\a$ were located at one particular nest, the
same conclusion~(\ref{prob2})  would be drawn about its distance from
particle $\b$. It may also be shown that this conclusion is unaltered
if we use the coordinates $\s_{\a k}$ rather than the coordinates
$r_{\a k}$.

\section{Physical properties - exclusion phenomena}
\setcounter{equation}{0}

In canonical quantum mechanics the Hamiltonian~(\ref{H})  corresponds to a system of
$n$ non-interacting oscillating particles, each of mass $m$. The allowed states of any
one particular particle are free and independent of the states of the other $n-1$
particles. For the Wigner quantum  oscillator this is not always the case. In fact
whenever $p<3n$, so that the corresponding irreducible representation of $sl(1|3n)$ is
atypical, constraints will apply. This is a consequence of the fact that if the sum of
any proper subset of the $\t_{\a i}$'s takes the value $p$ with $p<3n$, then the
remaining $\t_{\b j}$'s must all be zero. This implies an exclusion from certain states
even in the case $n=1$ of a single particle. For example, if $n=1$ and $p=2$ and
$\|p;\T\ra=\|2;1,1,\t_{13}\ra$ so that $\psi_{13}=\t_{11}+\t_{12}=2=p$, then
$\t_{13}=0$. It follows that while the state $\|2;1,1,0\ra$ is allowed, the state
$\|2;1,1,1\ra$ is forbidden. One might say that in the atypical case even a single
particle is not ``free'', or equivalently that the particle is ``excluded'' from being
in certain states.

This exclusion phenomenon is even more striking in the case of a multiparticle WQO with
$n>1$ and $p<3n$. This can be seen even in the simplest $n=2$ case of two particles,
for example when the order of the statistics  $p=3$, as in~(\ref{sl163}) and Table~2.
The relevant stationary states are given in the notation of~(\ref{basis}) by
$\|p;\T\ra=\|3;\t_{11},\t_{12},\t_{13},\t_{21},\t_{22},\t_{23}\ra$. If the first
particle is in the state $\t_{11}=\t_{12}=\t_{13}=1$, then since $q$ is constrained by
the condition $q\leq\min(p,3n)=\min(3,6)=3$ it follows that the second particle is
excluded from being in any state other than the state $\t_{21}=\t_{22}=\t_{23}=0$. This
state $\|3;1,1,1,0,0,0\ra$ is one of the $4$ states of $V_{gl(1)}^3\otimes V_{sl(3)}^0$
appearing in Table~2, having $p=q=3$, energy $E_3=3\hbar\omega/5$ and angular momentum
$M=0$. The contributions to the energy $E_q$ from each of the two particles may be
calculated by noting, quite generally from~(\ref{Halpha}),~(\ref{A+A-}) and~(\ref{Eq})
that \beq
  E_q=\sum_{\alpha=1}^n\ E_{\alpha,q}\quad\hbox{with}\quad
      E_{\alpha,q}={{\omega\hbar}\over{3n-1}}\sum_{i=1}^3 (p-q+\t_{\alpha i}).
\label{Ealpha} \eeq In the case of interest here for the state $\|3;1,1,1,0,0,0\ra$, we
have $n=2$, $p=q=3$, $\t_{11}=\t_{12}=\t_{13}=1$ and $\t_{21}=\t_{22}=\t_{23}=0$, so
that $E_{1,q}=3\hbar\omega/5$ and $E_{2,q}=0$. In addition, from~(\ref{Ma^2})  we have
\beq
   \hat{\M}_\a^2\|p;\T\ra = M^{(\a)}(M^{(\a)}+1)\|p;\T\ra
 \quad\hbox{with}\quad M^{(\a)}(M^{(\a)}+1)=\d_\a\ 2,
\label{Malpha} \eeq with $\d_\a$ defined by~(\ref{01}). For the state
$\|3;1,1,1,0,0,0\ra$ we have $\d_1=\d_2=0$, so that $M^{(1)}=M^{(2)}=0$. Finally
from~(\ref{rev}), for this same state with $p=q=3$ we have $r_{1,k}=\pm1$ and
$r_{2,k}=0$ for $k=1,2,3$.
Thus the particular state, $\t_{11}=\t_{12}=\t_{13}=1$, of the first particle, forces
the second to be such that $\t_{21}=\t_{22}=\t_{23}=0$ so that it sits at the origin
contributing no energy and no angular momentum to the total system. The same phenomenon
cannot occur for the CQO. In the notation of~(\ref{Phi})  if we have
$\|\Phi\ra=\|1,1,1,\phi_{21},\phi_{22},\phi_{23}\ra$ there is no restriction on the
parameters $\phi_{21},\phi_{22},\phi_{23}$ that determine the state of the second
particle.

The general conclusion in the multiparticle Wigner quantum oscillator case is
that despite the ``free'' nature of the Hamiltonian~(\ref{H}), the particles are
not always ``free''. They may ``interact'', in the sense that the state of a
particular particle may be constrained or even fixed by the states of the
other $n-1$ particles. The above $n=2$ and $p=3$ example illustrates this.
The interaction is of statistical origin, depending on the parameter $p$ in
our $A$-superstatistics model. This is very similar to the exclusion
statistics of Haldane~\cite{Haldane}, which plays an important role in condensed
matter physics.

As pointed out in Section~3, any comparison between the WQO and the CQO spectrum
of energy levels is preferably based on a comparison of the direct sum of an infinite
number of finite-dimensional irreducible representations of $sl(1|3n)$ and a single
infinite-dimensional irreducible representations of $osp(1|6n)$, that is,
in the notation of~(\ref{branch})  and~(\ref{osp}), a comparison of
\beq
      \sum_{p=0}^\infty V^p_{sl(1|3n)} \quad\quad\hbox{and}\quad\quad
     V^{\epsilon}_{osp(1|6n)}.
\label{SumVp}
\eeq
From~(\ref{Eq})  and~(\ref{BEq}) the complete sets of energy levels, indexed
by their level number $l$, are given in the two cases by
\beq
    {{l\,\hbar\omega}\over{(3n-1)}} \quad\quad\hbox{and}\quad\quad
   {{(3n+2l)\hbar\omega}\over{2}}
\quad\hbox{for}\quad l=0,1,2,\ldots \ .
\label{Elevels}
\eeq
In the WQO case the $l=0$ ground state has energy zero and the levels
are equally spaced with separation $\hbar\omega/(3n-1)$ which decreases
as the number of particles, $n$, increases. In the CQO case the $l=0$
ground state has energy $3n\hbar\omega/2$ and the levels are equally spaced
with separation $\hbar\omega$, independent of the number of particles $n$.

In both cases let the degeneracy of the $n$-particle level $l$ be denoted by
$d_{n,l}$ and the corresponding generating function be denoted by
\beq
     G_n(x)= \sum_{l=0}^\infty\ d_{n,l}\ x^l.
\label{Gen}
\eeq
Then for the WQO, using~(\ref{dimsl}), we find
\beq
    d_{n,l}= \cases{
        \quad 1& if $l=0$;\cr
        \quad 2& if $l\equiv 0\ (\mod 3n)$ and $l>0$;\cr
        {\ds{\left({{3n}\atop{r}}\right)}}&if $l\equiv r\ (\mod 3n)$ with $r>0$, \cr}
  \quad\hbox{and}\quad
    G_n(x)= {{(1+x)^{3n}}\over{(1-x^{3n})}}.
\label{DegenWQO}
\eeq
For the CQO the $n$-particle level $l$ corresponds to $q=l$ and its
degeneracy is given by~(\ref{dimslq}), so that
\beq
    d_{n,l}= \left({{3n-1+l}\atop{l}}\right)
\quad\hbox{and}\quad
    G_n(x)= {{1}\over{(1-x)^{3n}}}.
\label{DegenCQO}
\eeq
This degeneracy $d_{n,l}$ for the CQO increases without bound as $l$ increases
for any $n$, unlike the WQO case for which the degeneracy is bounded and in fact
periodic in $l$ for $l>0$.

\section{Concluding remarks}

To conclude, as promised in {\tt I}, we have taken the natural step of generalising a
one particle 3-dimensional WQO to an $n$-particle 3-dimensional WQO. The relevant
Fock space irreducible representations of $sl(1|3n)$, are as indicated previously,
of dimension $2^{3n}$ for the case of typical
representations, and less than this for atypical ones. With or without typicality
the energy levels are equally spaced, and we have shown how to
determine not just energy eigenstates but also mutual eigenstates of both
energy and angular momentum, as illustrated in detail in Table~2 for
the two particle case. One interesting feature of the $n$-particle model
is that for atypical representations there is an $A$-superstatistics
effect whereby one constituent particle may constrain the energy, angular
momentum and even configuration of another. In an extreme case with $p=3$
the existence of one particle in its lowest energy state forces all the
other particles to sit at the origin contributing no further energy or angular
momentum.

Moving to the $n$-particle case has also enabled us to explore in more detail the
sometimes unexpected consequences of the non-commutative geometry arising in this WQO model.
We find once again that the position of any one of the
particles may not be specified precisely. However, the possible results of measurements
of various coordinates lead us to identify various sites or nests whose precise positions
turn out to be a function of the coordinates we choose to measure. The distances between
these nests associated with different particles are not in fact what one might have expected,
namely the
square roots of the eigenvalues of the operators ${\hat{d}}^2_{\a\b}(t)$ associated with
the square of the distance between any two particles
specified by $\a$ and $\b$. Instead, as in more conventional quantum theory
models, the expectation value ${\overline{{\hat{d}}^2}}_{\a\b}(t)$ of
the square of the distance operator ${\hat{d}}^2_{\a\b}(t)$ in each stationary state
$\|p;\T\ra$ gives the average square distance between the various nests of the two particles
associated with the stationary state $\|p;\T\ra$, with each nest occupied with various
possible probabilities.

Consideration of appropriate infinite-dimensional representations has enabled us to
compare and contrast the classical canonical quantum oscillator and our non-standard
Wigner quantum oscillator. The latter has the unusual feature that the equally spaced
energy levels become closer together, but remain equally spaced as the number of
particles is increased. In addition their degeneracy remains bounded and, above the
ground state, the degeneracy is periodic in the level number for any fixed number of
particles.

\bigskip
\noindent{\bf Acknowledgements}
\medskip

\noindent
The work was supported by a Royal Society Joint Project Grant between the UK
and Bulgaria. In addition NIS was supported by a Marie Curie Individual Fellowship of
the European Community Programme `Improving the Human Research Potential and the
Socio-Economic Knowledge Base' under contract number HPMF-CT-2002-01571.

\end{document}